\documentclass[useAMS,usenatbib]{mn2e}
\usepackage[totalwidth=515pt,totalheight=660pt,left=1.4cm,right=1.4cm]{geometry}
\usepackage{graphicx,amssymb,color}
\usepackage[normalem]{ulem}
\input psfig

\title[Forming WD discs I: tidal disruption]
{Formation of planetary debris discs around white dwarfs I:
Tidal disruption of an extremely eccentric asteroid}
\author[Veras, Leinhardt, Bonsor \& G\"{a}nsicke]{
Dimitri Veras$^{1}$\thanks{E-mail:d.veras@warwick.ac.uk},
Zo\"{e} M. Leinhardt$^{2}$,
Amy Bonsor$^{2}$,
Boris T. G\"{a}nsicke$^{1}$
\\
$^{1}$Department of Physics, University of Warwick, Coventry CV4 7AL, UK
\\
$^{2}$The School of Physics, University of Bristol, Bristol BS8 1TL, UK
}

\begin{document}

\date{Accepted 2014 September 07. Received 2014 September 07; in original form 2014 June 20}

\pagerange{\pageref{firstpage}--\pageref{lastpage}} \pubyear{XXXX} 
%\onecolumn

\maketitle

\label{firstpage}

\begin{abstract}
25\%-50\% of all white dwarfs (WDs) host observable and dynamically
active remnant planetary systems based on the presence of close-in 
circumstellar dust and gas and photospheric metal pollution.
Currently-accepted theoretical explanations for the origin of this matter include 
asteroids that survive the star's giant branch evolution 
at au-scale distances and are subsequently perturbed onto WD-grazing orbits 
following stellar mass loss. 
In this work we investigate the tidal disruption of these 
highly-eccentric ($e > 0.98$) asteroids as they approach 
and tidally disrupt around the WD. 
We analytically compute the disruption 
timescale and compare the result with fully self-consistent numerical simulations of
rubble piles by using the $N$-body code {\tt PKDGRAV}.  We find that this timescale
is highly dependent on the orbit's pericentre and largely independent of its semi-major axis. 
We establish that spherical asteroids readily break up and 
form {\it highly eccentric collisionless rings}, which do not accrete 
onto the WD without additional forces such as radiation or sublimation. This finding highlights the critical importance of such forces in the physics of WD planetary systems.
\end{abstract}


\section{Introduction}

The realisation that the rocky material which pollutes white dwarf (WD) atmospheres 
primarily originates
from circumstellar debris and not the interstellar medium 
\citep{kilred2007,gaeetal2008,faretal2009,juretal2009,faretal2010} has revolutionised the study of evolved planetary systems.  Precise and extensive observations of metal abundances
in WD atmospheres \citep{zucetal2003,zucetal2010,koeetal2014} suggest the presence of dynamically-active systems.  
This notion is reinforced by secure observations of orbiting dust 
\citep{zucbec1987,faretal2012,xujur2012} and gas \citep{gaeetal2006,gaeetal2007,gaeetal2008,debetal2012a}.  

Whereas the presence of orbiting dust, and its approximate radial
distribution, is inferred from measurements of infrared excess
luminosity, the existence of gaseous material is instead inferred from
metal emission lines, in particular the Ca{\small II} triplet
near 8600\,\AA. The morphology of the emission line profiles reflect
the velocity field of the gas in motion around the WD \citep[see][]{hormar1986}, 
and thereby provide insight into the
spatial distribution of the gas. All of the gaseous system signatures
so far discovered constrain the location of the matter to be within or
around the WD tidal disruption radius, at about one Solar
radius. The two best-studied systems exhibit a noticeable asymmetry in
the shape of the double-peaked line profiles, suggesting
eccentricities in the range 0.02 to 0.2 \citep{gaeetal2006,gaeetal2008}. 
In addition, in at least two cases, the shape and strength of
the emission lines vary between observations obtained a few years
apart, demonstrating evolution of at least the gaseous component of
the disc structure on relatively short timescales (Fig. 3 of 
\citeauthor{gaeetal2008} \citeyear{gaeetal2008}, Wilson et al. 2014 In Prep).

The complex structure of the gaseous material highlights
the dangers of, and simply prove incorrect, assuming all
material disrupted around the WD forms a circular disc.
What is clear is that for multiple systems, the material is within
or around the tidal disruption radius.  Hence, these structures cannot have formed during earlier
stellar phases, because otherwise they would have resided inside of 
the progenitor!  How they formed during the WD phase remains an outstanding question.  

As a first step towards finding an answer, in this work we consider the disruption process 
of a rubble-pile asteroid around a WD with help from the sophisticated $N$-body numerical code
{\tt PKDGRAV} \citep{ricetal2000,stadel2001}.  Although the tidal breakup of rocky asteroidal material has previously been proposed \citep{graetal1990,jura2003,beasok2013}, the progenitors of these discs could be comets, moons or planets.  
However, recent theoretical work has favoured asteroids \citep{bonetal2011,debetal2012b,frehan2014} 
primarily due to the low frequency of planetary collisions with WDs \citep{veretal2013,musetal2014}
and the compositional inconsistencies \citep{zucetal2007} and dynamical difficulties \citep{stoetal2014,veretal2014a} 
of comet accretion; investigation of moons is needed.  Henceforth we use the term {\it asteroids} to refer to any 
small bodies.

In the Solar System, asteroids are known to reside within several tens of au of the Sun.  Exo-asteroids
at similar separations which survive dynamical instabilities or tidal engulfment during the giant 
branch phases of their 
parent stars will harbour wider orbits by a factor of a few, and not be ejected
due to mass loss alone \citep{veretal2011} even for particularly violent stellar structure
assumptions \citep{verwya2012}.  At a minimum, the asteroids need to reside beyond about a couple of
au to avoid engulfment into their star's giant branch envelope \citep{musvil2012}.  Therefore
remnant asteroids are expected to reside at distances between a few au and a couple hundred au, and these asteroids
can be flung towards the WD only on extremely eccentric orbits.  Here we are unconcerned with the 
dynamical architectures that would be necessary to propel the asteroid to the WD in this manner 
(see \citealt*{bonetal2011,debetal2012b,frehan2014}) but rather focus entirely on the 
disruption process.

This paper contains 6 sections, and introduces several timescales, which are summarised 
in Table 1 for ease of reference.  
Section 2 establishes the location of the critical disruption sphere
and provides a link to the WD radius and mass.  Section 3 describes the orbit of an extremely
eccentric asteroid with respect to the critical disruption sphere.  In Section 4, we analytically 
determine the timescale for forming an eccentric ring from the disruption.  We set up and run our 
numerical simulations of the disruption in Section 5 before concluding in Section 6.

\begin{table}
 \centering
  \caption{Timescales used in this paper.}
  \begin{tabular}{@{}ccccc@{}}
  \hline
 Timescale & Timescale & Equation    \\
 Symbol & Name & Number(s) \\
 \hline
$t_{\rm c}$       & Crossing or Disruption  & \ref{cross1}, \ref{newtc}, \ref{limtc}  \\
$P_{\omega}$      & GR Pericentre Precession  & \ref{GRinf}   \\
$t_{\rm fill}$    & Filling  & \ref{fill1}   \\
$\tau_{\rm dyn} $ & Dynamical & \ref{tdyn}   \\
$\tau_{\rm orb} $ & Orbital & \ref{torb}  \\
$\tau_{\rm enc} $ & Encounter & \ref{tenc1}-\ref{tenc2}   \\
$\Delta t$      & Simulation Timestep & \ref{timestep} \\
\hline
\end{tabular}
\end{table}

\section{Critical disruption radius}

We define disruption simply as a significant morphological change.  
Both observations and theory provide strong insight into the critical value at which 
disruption will occur.  Nearly all known planetary rings, which may have been formed from 
the disruption of asteroids, are observed to orbit within a few planetary radii from the 
centre of the planets.  The rings around the Centaur Chariklo also lie within a few asteroid 
radii from the centre of the asteroid \citep[Section 7 of the supplement of][]{braetal2014}. Force 
balance arguments \citep[e.g. pgs. 158-159 of][]{murder1999} demonstrate that the critical 
disruption radius $r_{\rm c}$ has the dependencies given by 

\begin{equation}
r_{\rm c} \propto \left( \frac{M_{\rm WD}}{M} \right)^{1/3} R
,
\end{equation}

\noindent{}where $M_{\rm WD}$ and $M$ are the masses of the WD and asteroid and $R$ is 
a fiducial radius of the (not necessarily spherical) asteroid.

The proportionality constant is model-dependent and is based on the shapes, compositions, 
spin states, orbital states, and criteria used for disruption.  
The constant may include functions of the tensile or shear strengths of the asteroid 
\citep[e.g.][]{davidsson1999}, and may be determined from numerical simulations rather 
than analytical considerations in order to be liberated from the assumptions of the latter 
\citep{ricetal1998}.  The constant will change depending on whether the asteroid is modelled 
to simply crack, deform, or dissociate entirely.  The disruption radius is famously named 
after Edouard Roche, although his pioneering calculation was based on just a single set of 
assumptions.

Because tidal disruption is a dynamic process that in reality cannot be reduced to a simple 
critical radius criterion, our model makes implicit assumptions.  These are that
the asteroid is roughly spherical, frictionless and not rotating, or rotating synchronously.  These assumptions
may both overestimate and underestimate the disruption radius 
\citep[e.g.][]{sritre1992,aspben1994,aspben1996,botetal1997,ricetal1998,movetal2012}.
Hence, our proceeding treatment is an oversimplification.  However, in order to
obtain analytical results, we adopt these assumptions for the remainder of the manuscript.

For our purposes, a useful expression of the disruption radius is 

\begin{equation}
\frac{r_{\rm c}}{R_{\odot}} = C \left( \frac{M_{\rm WD}}{0.6 M_{\odot}} \right)^{1/3} \left( \frac{\rho}{3 \ {\rm g/cm^3}} \right)^{-1/3}
\label{bearr}
\end{equation}

\noindent{where} $C$ is a constant ranging from about $0.85$ to $1.89$ \citep{beasok2013},
and $\rho$ is the assumed density of the asteroid.   
The value of $0.6 M_{\odot}$ may be considered as a fiducial WD mass given the mass
distribution of all observed WDs \citep{lieetal2005,camenzind2007,faletal2010,treetal2013}.
Due to observational evidence that the vast majority of asteroids have densities which satisfy 
$\rho \gtrsim 1$~g/cm$^3$ \citep[Table 1 of][]{carry2012}, we find 

\begin{equation}
{\rm max}\left[r_{\rm c}\left(M_{\rm WD}\right)\right] 
\equiv
r^\mathrm{max}_\mathrm{c}(M_\mathrm{WD})
\approx 
2.73 \left( \frac{M_{\rm WD}}{0.6 M_{\odot}} \right)^{1/3} R_{\odot}
\label{maxrc1}
\end{equation}

\noindent{}where we have assumed the maximum value of $C$ and minimum value of $\rho$.  

Further, by invoking the 
Chandrasekhar Limit, which gives the maximum WD mass ($\equiv M_\mathrm{Ch} = 1.4M_{\odot}$), the 
maximum value of 
$r^\mathrm{max}_\mathrm{c}(M_\mathrm{WD})$ is $r^\mathrm{max}_\mathrm{c}(M_\mathrm{Ch}) 
= 3.6 R_{\odot} = 0.017 \ {\rm au} = 2.5 \times 10^6$ km.  This value is at least a few 
hundred times greater than the radius of the WD, which, for a typical 
WD mass of $0.6 M_{\odot}$, is $\simeq 0.015 R_{\odot}$ 
\citep{hamsal1961,holetal2012,paretal2012}.
Equation \ref{maxrc1} usefully demonstrates just how small the disruption region is.  
Any asteroid whose disruption we wish to model must eventually pass inside a sphere with 
a radius of $r_{\rm c}$ centred on the WD.  Also, because the asteroid might collide with the
WD, we must compute $R_{\rm WD}$.

Both observations and theory demonstrate that mass alone does not uniquely determine the extent of this surface; 
temperature is another dependence \citep[e.g.][]{panetal2000}. If we neglect this temperature dependence, then 
equations (27-28) of \cite{nauenberg1972} 
link WD mass and radius through the following relation

\begin{equation}
\frac{R_{\rm WD}}{R_{\odot}}
\approx
0.0127 
\left( \frac{M_{\rm WD}}{M_{\odot}} \right)^{-1/3}
\sqrt{
1 - 0.607
\left( \frac{M_{\rm WD}}{M_{\odot}} \right)^{4/3}
}
,
\label{MR}
\end{equation}

\noindent{}where we have assumed a mean molecular weight per electron of 
$2$ \citep{hamsal1961}.  
\noindent{}Another popular relation is from 
equation 15 of \cite{verrap1988}.

\[
\frac{R_{\rm WD}}{R_{\odot}}
\approx
0.0114
\sqrt{
\left( \frac{M_{\rm WD}}{1.44 M_{\odot}} \right)^{-2/3}
-
\left( \frac{M_{\rm WD}}{1.44 M_{\odot}} \right)^{2/3}
}
\]

\begin{equation}
\times
\left[
1 + 3.5 
\left( \frac{M_{\rm WD}}{0.00057 M_{\odot}} \right)^{-2/3}
+
\left( \frac{M_{\rm WD}}{0.00057 M_{\odot}} \right)^{-1}
\right]^{-2/3}
.
\label{MR2}
\end{equation}

\noindent{}Both relations produce nearly identical results, with a variation of just a few percent.
Further, both relations reproduce well the latest observational results within the error bars
\citep[Fig. 4 of][]{bouetal2014}.

The higher the WD mass, the smaller the WD radius, so that  
for $M_\mathrm{WD}<M_\mathrm{Ch}$,
$R_{\rm WD} > 6.4 \times 10^{-4} R_{\odot} \approx 3.0 \times 10^{-6}$au $\approx 445$ 
km\footnote{For perspective, this extreme WD would rank 5th in size amongst the
Uranian satellites.}.  A small number of WDs with very low masses, down to 
$~0.17 M_{\odot}$ \citep[e.g.][]{broetal2013,heretal2013}  
have been discovered. They are all products of close binary interactions, and it is currently 
not clear if these systems have any relevance in the context of evolved planetary systems; 
we no longer consider low mass WDs for the remainder of the paper.
For our calculations, we adopt
the canonical $0.6 M_{\odot}$ mass, which corresponds
to a value of $R_{\rm WD} = 0.0126 R_{\odot} \approx 8750$ km from equation (\ref{MR})
\footnote{This WD would be just 37 per cent larger in radius than the Earth.}.
Nevertheless, we retain WD mass in all our formulae for future applications.
In order to express the critical disruption radius in terms of $R_{\rm WD}$,
we may combine equation (\ref{MR}) with equation (\ref{bearr}). 
Overall, these relations show that disruption 
predominately occurs at a distance of $\sim 10^5-10^6$ km 
($7 \times 10^{-4} - 7 \times 10^{-3}$ au) from the centre of the WD.

\section{Orbit characteristics}

Now that we have quantified the region that asteroids must pass through for disruption 
to occur, we consider the asteroid orbits themselves.  The orbits are noteworthy
because of their extreme eccentricity.  In fact, any asteroid with a semimajor 
axis $a > 1$ au (like the vast majority of Solar System asteroids) must have an extremely 
eccentric ($e > 0.983$) orbit in order to achieve a pericentre within the maximum possible 
disruption radius [$r^\mathrm{max}_\mathrm{c}(M_\mathrm{Ch})$].
This section will provide detailed characteristics of this orbit, particularly when the 
asteroid is within the disruption sphere.

\subsection{The speed at pericentre}

The speed of the asteroid at pericentre, $v_{q}$, is remarkably high.
If the semimajor axis and pericentre of the asteroid's orbit are denoted by $a$ and $q$, then

\begin{eqnarray}
v_{q} &\approx& 23.1 \frac{\rm km}{\rm s} \left( \frac{M_{\rm WD}}{0.6 M_{\odot}} \right)^{1/2}
                                 \left( \frac{a}{1 \ {\rm au}} \right)^{-1/2}
                                 \left(\frac{1 + e}{1 - e}\right)^{1/2}
\label{vpfora}
\\
&=& 730 \frac{\rm km}{\rm s} \left( \frac{M_{\rm WD}}{0.6 M_{\odot}} \right)^{1/2}
                   \left( \frac{q}{0.001 \ {\rm au}} \right)^{-1/2}
                   \sqrt{1+e}
.
\end{eqnarray}

\subsection{The range of interesting pericentres}

Disruption can occur only when the pericentre is within $r_{\rm c}$.  This fact,
along with our previous findings, allows us to quantify the minimum ($q_{\rm min}$) and 
maximum ($q_{\rm max}$) pericentres we will consider in this paper.  Both $q_{\rm min}$ and 
$q_{\rm max}$ are expressed in terms of radius and mass of the WD through equations 
(\ref{MR}) and (\ref{maxrc1}), respectively.  Consequently,

\begin{eqnarray}
&&\frac{q_{\rm max}}{q_{\rm min}}
= 
\frac{r^\mathrm{max}_\mathrm{c}(M_\mathrm{WD})}{R_{\rm WD}}
\nonumber
\\
&&\approx 254 \left(\frac{M_{\rm WD}}{M_{\odot}}\right)^{2/3}
\left[
1 - 0.607
\left( \frac{M_{\rm WD}}{M_{\odot}} \right)^{4/3}
\right]^{-1/2}
,
\label{qmaxqmin}
\end{eqnarray}

\noindent{}which yields a value of 217 for a WD mass of $0.6 M_{\odot}$.

\subsection{Entering and exiting disruption sphere}

The part of an asteroid's orbit of the greatest interest, and the part which we will model 
numerically, is the region inside of the disruption sphere.  Suppose the given orbit is centred on 
a Cartesian reference grid such that the star lies at the fixed position $(ae,0)$.  
Without loss of generality, assume the asteroid moves counterclockwise. 
Then the entry and exit points of the disruption sphere along the orbit, assuming the orbit remains static through pericentre passage, are

\begin{equation}
\left(x_{\rm e}, y_{\rm e}  \right) = 
\left( \frac{a - r_{\rm c}}{e},  
\pm \frac{1}{e} \sqrt{\left(1 - e^2\right) \left[2 a r_{\rm c} - r_{\rm c}^2 - a^2 \left(1 - e^2\right) \right]}
\right)
.
\label{xeye}
\end{equation}

The entry and exit distance, $r_{\rm e}$, from the star is just $r_{\rm e} = r_{\rm c}$, and 
the speed at these entry and exit points, $v_{\rm e}$, is

\begin{eqnarray}
v_{\rm e} &\approx& \sqrt{GM_{\rm WD} \left(\frac{2}{r_{\rm e}}  -\frac{1}{a} \right)}
\nonumber
\\
&=& 23.1 \frac{\rm km}{\rm s} \left( \frac{M_{\rm WD}}{0.6 M_{\odot}} \right)^{1/2}
                                 \left( \frac{a}{1 \ {\rm au}} \right)^{-1/2}
                                 \left(\frac{2 + e}{2 - e}\right)^{1/2}.
\label{vefora}
\end{eqnarray}

\noindent{}Equation (\ref{vefora}) should be compared with equation (\ref{vpfora}).  One then observes in the limit of $e \rightarrow 1$, for $e> 0.983$, we obtain

\begin{equation}
\frac{v_{\rm e}}{v_q} \approx  \sqrt{\frac{3}{2}\left(1 - e\right)} < 16\%
\label{vevp}
\end{equation}

\noindent{meaning} that the pericentre velocity typically exceeds both the entry and exit velocity by about one order of magnitude.
This result showcases how drastically the asteroid's velocity changes just within the small disruption sphere even if no disruption occurs.  

\subsection{Time spent within the disruption sphere}

The time spent within the disruption sphere, $t_{\rm c}$, will help us predict how the extent of disruption is linked to a particular orbit.  We estimate this crossing timescale by assuming the orbit remains static.  We obtain

\begin{equation}
t_{\rm c} = \frac{2 \left|\Pi_{\rm e}\right|}{n}
\label{cross1}
\end{equation}

\noindent{}where the mean motion, $n \approx \sqrt{GM_{\rm WD}/a^3}$ (excluding the relatively tiny mass of the asteroid), and the 
mean anomaly, $\Pi_{\rm e}$, at either the entry or exit point, is given by Kepler's equation ($\Pi_{\rm e} = E_{\rm e} - e\sin{E_{\rm e}} $).
The eccentric anomaly, $E_{\rm e}$, at these points is obtained from

\begin{equation}
\cos{E_{\rm e}} = \frac{1}{e} \left(1 - \frac{r_{\rm c}}{a} \right)
.
\label{cosEe}
\end{equation}

\noindent{}Knowledge of $t_{\rm c}$ is particularly important in order to effectively set up numerical simulations.

Because the asteroid must not hit the WD and be within the disruption sphere, we have $R_{\rm WD} < q < r_c$.
The time spent in the disruption sphere is equal to zero for $q = r_{c}$ and varies as $q$ approaches $R_{\rm WD}$,
when the asteroid would be moving fastest (equation \ref{vpfora}).
We can compute the maximum time by first rewriting $t_c$ as

\begin{eqnarray}
t_{\rm c} &=& 2\sqrt{\frac{a^3}{GM_{\rm WD}}}
\nonumber
\\
&\times&
\left\lbrace
\cos^{-1}\left[ \frac{1 - \frac{r_c}{a} }{1 - \frac{q}{a} } \right]
-
\sqrt{ \left(1 - \frac{q}{a} \right)^2 - \left(1 - \frac{r_c}{a} \right)^2}
\right\rbrace
.
\label{newtc}
\end{eqnarray}

\noindent{}Consequently, the value of $q$ which gives the maximum $t_{\rm c}$ is

\begin{equation}
q' = a \left[ 1 - \sqrt{1 - \frac{r_c}{a} } \right]
\label{qprime}
\end{equation}

\noindent{}and

\begin{equation}
{\rm max}\left(t_{\rm c}\right) 
= 
t_{\rm c}\left(q'\right)
=
\sin^{-1}\left(\frac{r_c}{a}\right)
-
\sqrt{\frac{r_c}{a} \left(1 - \frac{r_c}{a} \right)  }
.
\end{equation}

\begin{figure}
\centerline{
\psfig{figure=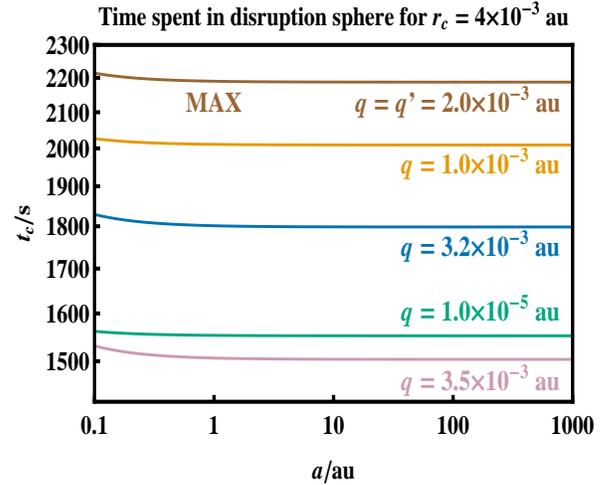,height=6.5cm,width=8cm}
}
\caption{Time an asteroid spends per orbit in a spherical region around a $0.6 M_{\odot}$ WD where disruption can occur.  The value of $t_c$ is crucially dependent upon the pericentre $q$ but almost independent of the semimajor axis $a$ of the asteroid's orbit.  These curves assume a disruption sphere radius of $r_{\rm c} = 4 \times 10^{-3}$ au.  The top curve shows the peak value of the disruption time (equation \ref{limtcqc}).}
\label{tc1}
\end{figure}

\noindent{}Figure \ref{tc1} quantifies these equations for a fiducial WD with $M_{\rm WD} = 0.6M_{\odot}$.  
Note how steeply the time spent in the disruption sphere decreases as a function of $q$ after the maximum value 
is attained at $q'$.  The plot demonstrates that for a fixed value of $q$, the time spent in the disruption 
sphere is nearly independent of $a$ except for $a \ll 0.1$ au.  Importantly then, we expect the disruption
characteristics to be independent of $a$ for all semimajor axes which could survive engulfment on the giant branch phases of stellar evolution.  The value of $t_{\rm c}$ which satisfies nearly all relevant 
values of $a$ is

\begin{equation}
\lim_{a \rightarrow \infty}{\left[t_{\rm c}\right]} =
\frac{2\sqrt{2}}
{3 \sqrt{GM_{\rm WD}}}
\left[
\frac
{r_{c}^2 + q r_c - 2q^2}
{\sqrt{r_c - q}}
\right]
\label{limtc}
\end{equation}

\noindent{}with the maximum of these values occurring at

\begin{equation}
\lim_{a \rightarrow \infty}{\left[t_{\rm c}\left(q'\right)\right]} =
\frac{4}{3}
\sqrt{ 
\frac{r_{c}^3}{GM_{\rm WD}}  
}
,
\label{limtcqc}
\end{equation}

\noindent{}which is a factor of $3/(2\pi)$ times the orbital period of an object that
travels along the disruption boundary.

\begin{figure}
\centerline{
\psfig{figure=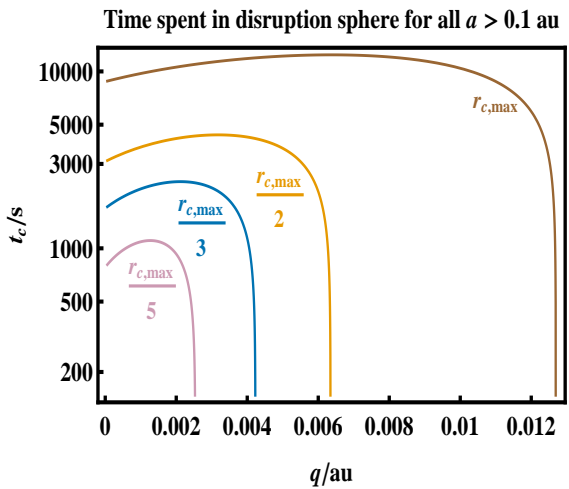,height=6.5cm,width=8cm}
}
\caption{Like Fig. \ref{tc1}, except the crossing time here is plotted as a function of the pericentre $q$ for four different
disruption spheres with radii $r_{\rm c}$.  The maximum possible value of $r_{\rm c}$ for a $0.6 M_{\odot}$ star is 
${\rm max}\left[r_{\rm c}\right] \approx 2.73 R_{\odot}$ ($0.013$ au), and arises from equations (\ref{bearr})-(\ref{maxrc1}),
where the unknown parameters are $C$ and $\rho$.  
These curves do not visibly change when $a$ is varied beyond about 0.1 au.
The peak of each curve occurs approximately halfway between the centre and edge of the sphere, as can be deduced by Taylor
expanding equation (\ref{qprime}) about small values of $r_{\rm c}/a$.
}
\label{tc2}
\end{figure}

Now we consider the distribution of $t_{\rm c}$ within the sphere as a function of $q$.
Figure \ref{tc2} illustrates the result, and that the maximum crossing time does not occur
at the WD surface.  The reason is because the asteroid is moving the fastest when skimming the surface.

Instead, the maximum disruption or crossing time occurs at $q \approx r_{\rm c}/2$, which is equivalent to 
the result of Taylor expanding equation (\ref{qprime}) about small values of $r_{\rm c}/a$.  
Note the asymmetry in the curves; the disruption timescale always exceeds $70\%$ of the 
maximum value unless $q \gtrsim 0.87 r_{\rm c}$.  Therefore, as long as the pericentre is not close to the edge 
of the disruption sphere, the disruption crossing time is approximately constant.  In conclusion, the time 
spent within the disruption radius is typically a few 1000s.
 
Within the disruption radius, the internal changes the asteroid undergoes are complex and may be
strongly dependent on our assumptions of sphericity, frictionlessness and no spin.  For example, as
observed by \cite{movetal2012}, the size distribution of the granular constituents of a real
asteroid will affect the relationship between confining pressure and the maximum allowed shear
stress.  Consequently, disruption may occur within a region other than a sphere, in which case
our value of $t_{\rm c}$ would have to be modified.  Further, the shape of the disruption region might change as
the asteroid is passing through and changing its own shape and/or spin.  Regardless, as illustrated 
by equation (\ref{newtc}) and Figure \ref{tc1},
the total time spent at a pericentre passage is largely independent of the orbit's semimajor
axis.  This result is independent of the detailed internal dynamical interaction which occurs
at the pericentre.

%%%%%%%%%%%%%%%% Figure
%%%%%%%%%%%%%%%% Figure

\subsection{Contribution from general relativity}

As the asteroid approaches the WD, the star will curve spacetime, thereby altering the trajectory of
the asteroid from the Newtonian value.  Here we evaluate this contribution, showing it to be negligible along individual orbits but not necessarily so over secular timescales.

For a single nearly parabolic orbit, 
\cite{veras2014} showed that the maximum deviation at the
pericentre of a $0.6 M_{\odot}$ star is approximately equal to 2.6 km.  
When compared with the radius of our adopted WD (8750 km), the 
extent of the disruption sphere ($10^5$ - $10^6$ km), and the error induced by taking the limit
of large $a$ in computations (see Fig. \ref{tc1}), this correction is negligible.  However, he 
points out that as $q$ remains fixed as $a$ increases, the approximation get worse, such that
for $a = 10^5$ au and $q = 0.1$ au, the error in the estimation is of order unity.  Nevertheless, 
this error is still negligible, and that situation is more relevant for long-period comets than 
for asteroids.

Over many orbits, general relativity will torque the asteroid's 
argument (or longitude) of pericentre.  This angle will precess
over one complete orbit in a time $P_{\omega}$, where

\begin{equation}
P_{\omega} \approx 0.15 \ {\rm Myr}
\left[
\frac{1 - e^2}
{1-0.999^2}
\right]
\left(
\frac{M_{\rm WD}}{0.6 M_{\odot}}
\right)^{-3/2}
\left(
\frac{a}{1 \ {\rm au}}
\right)^{5/2}
.
\label{GRinf}
\end{equation}

\noindent{}Therefore, a precession of a few degrees may occur on thousand-year
timescales, which correspond to a few tens of orbits for sufficiently far-away
progenitors.  Consequently, general relativity would enhance the possibility
of collisions amongst debris which is flung out to different semimajor axes.
We will consider this possibility, particularly with the disruption of multiple
asteroids, in future work.

%%%%%%%%%%%%%%%% Figure
%%%%%%%%%%%%%%%% Figure

\section{Eccentric ring formation timescale}

Before performing numerical simulations, we can make theoretical predictions about 
disrupted debris.  In particular, we predict that the debris will form an eccentric
ring which
follows the original orbit.  Below we estimate the timescale for formation of this
eccentric ring.  The extent of the agreement with numerical integrations will help
investigators determine the usability of the theoretical model in future studies.

Our treatment follows the formulation presented by \cite{hahret1998}, which agreed well
with numerical simulations of the disruption of comet Shoemaker-Levy 9.  Here, suppose
our asteroid is composed of many point mass particles.  Later, in our numerical simulations
(next section), these particles will adopt nonzero radii.  Let all variables with subscript
``P'' refer to a specific but arbitrary particle.  Variables without subscripts refer to
the asteroid.  In what follows, assume that the breakup
is instantaneous and occurs at $r_{\rm b}$, and that the particles evolve independently of each other
(are {\it collisionless}) immediately after the breakup.  Our formulation is 
independent of both $r_{\rm e}$ and $r_{\rm c}$.

All particles will move with the same velocity before the asteroid breaks up.  Hence,

\begin{equation}
v_{\rm P}^2 = v^2 = G \left(M_{\rm WD} + M\right)
\left(\frac{2}{r_{\rm b}} - \frac{1}{a}\right)
.
\end{equation}

\noindent{}By conservation of energy,

\begin{equation}
-\frac{G M_{\rm WD} M_{\rm P}}{2 a_{\rm P}} = \frac{1}{2} M_{\rm P} v_{\rm P}^2 
- \frac{G M_{\rm WD} M_{\rm P}}{r_{\rm P}}
\end{equation}

\noindent{}which gives

\begin{equation}
a_{\rm P} = \frac{a r_{\rm b} r_{\rm P}}
{\left(\frac{M}{M_{\rm WD}}\right) r_{\rm P} \left(r_{\rm b} - 2a\right) + 2a \left(r_{\rm b} - r_{\rm P}\right) 
+ r_{\rm b} r_{\rm P}}
.
\label{apring}
\end{equation}

When $r_{\rm P} = r_{\rm b}$, that particle continues along the asteroid's original elliptic
orbit.  When $r_{\rm P} < r_{\rm b}$, that particle will have an elliptical orbit.  When
$r_{\rm P} > r_{\rm b}$, that particle can harbour an elliptical, parabolic or hyperbolic
orbit.  For this last case, consider equation (\ref{apring}).
Properties of conic sections dictate that the distance at which 
the particle's orbit becomes parabolic is

\begin{equation}
r_{\rm crit} = \frac{2ar_{\rm b}}{\left(1 + \frac{M}{M_{\rm WD}} \right) \left(2a - r_{\rm b}\right)}
\approx
\frac{2ar_{\rm b}}{2a - r_{\rm b}}
\end{equation} 

\noindent{}such that the particle's orbit remains elliptical if $r_{\rm P} < r_{\rm crit}$
or becomes hyperbolic if $r_{\rm P} > r_{\rm crit}$.  Note that for asteroids around WDs,
we can assume $M/M_{\rm WD} \approx 0$ because that 
ratio is about 10 orders of magnitude smaller than any ratio of relevant length scales in this problem.

The velocity gradient between bound debris will fill out a ring.  The initial spatial
distance between the bound debris will determine the formation timescale.  This distance
can be up to the entire asteroid diameter, or at minimum the asteroid radius, as all particles between
the asteroid centre and the closest point to the WD must remain on bound orbits. The value
of $r_{\rm crit}$ determines whether or not all of the particles will remain on bound orbits.
Consequently, the debris will fill an entire orbit in a time

\begin{equation}
t_{\rm fill} = \frac{2\pi}{n\left(r_p = r_{\rm b}-R\right) - n\left(r_p = r_{\rm b}+{\rm min}(r_{\rm crit}-r_{\rm b}, R)\right)} 
.
\label{tfill1}
\end{equation}

\noindent{}If $t_{\rm fill}$ is expressed in terms of the asteroid's (original) orbital period ($T$), then 
we finally obtain

\[
\frac{t_{\rm fill}}{T} = 
\frac{n\left(r_p = r_{\rm b}\right)}{n\left(r_p = r_{\rm b}-R\right) - n\left(r_p = r_{\rm b}+{\rm min}(r_{\rm crit}-r_{\rm b}, R)\right)} 
\nonumber
\]

\begin{eqnarray}
&=&
r_{\rm b}^{\frac{3}{2}}
\bigg[  
\bigg\lbrace 
\frac
{r_{\rm b}^2 + 2aR - r_{\rm b}R}
{r_{\rm b}-R}
\bigg\rbrace^{\frac{3}{2}}
\nonumber
\\
&-&
\bigg\lbrace 
\frac
{r_{\rm b}^2 - 2a\times{\rm min}(r_{\rm crit}-r_{\rm b}, R) + r_{\rm b}{\rm min}(r_{\rm crit}-r_{\rm b}, R)}
{r_{\rm b}+{\rm min}(r_{\rm crit}-r_{\rm b}, R)}
\bigg\rbrace^{\frac{3}{2}}
\bigg]^{-1}
.
\nonumber
\\
\label{fill1}
\end{eqnarray}

\begin{figure}
\centerline{
\psfig{figure=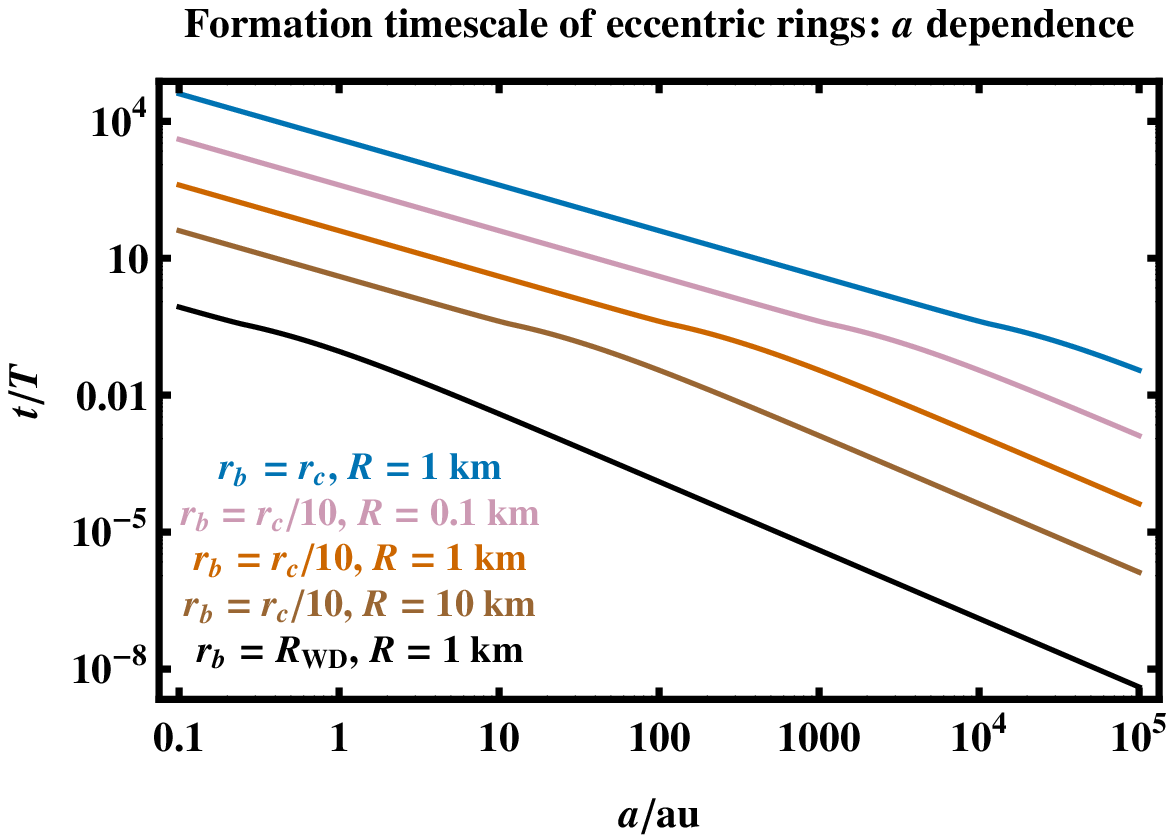,height=6.5cm,width=8cm}
}
\centerline{
\psfig{figure=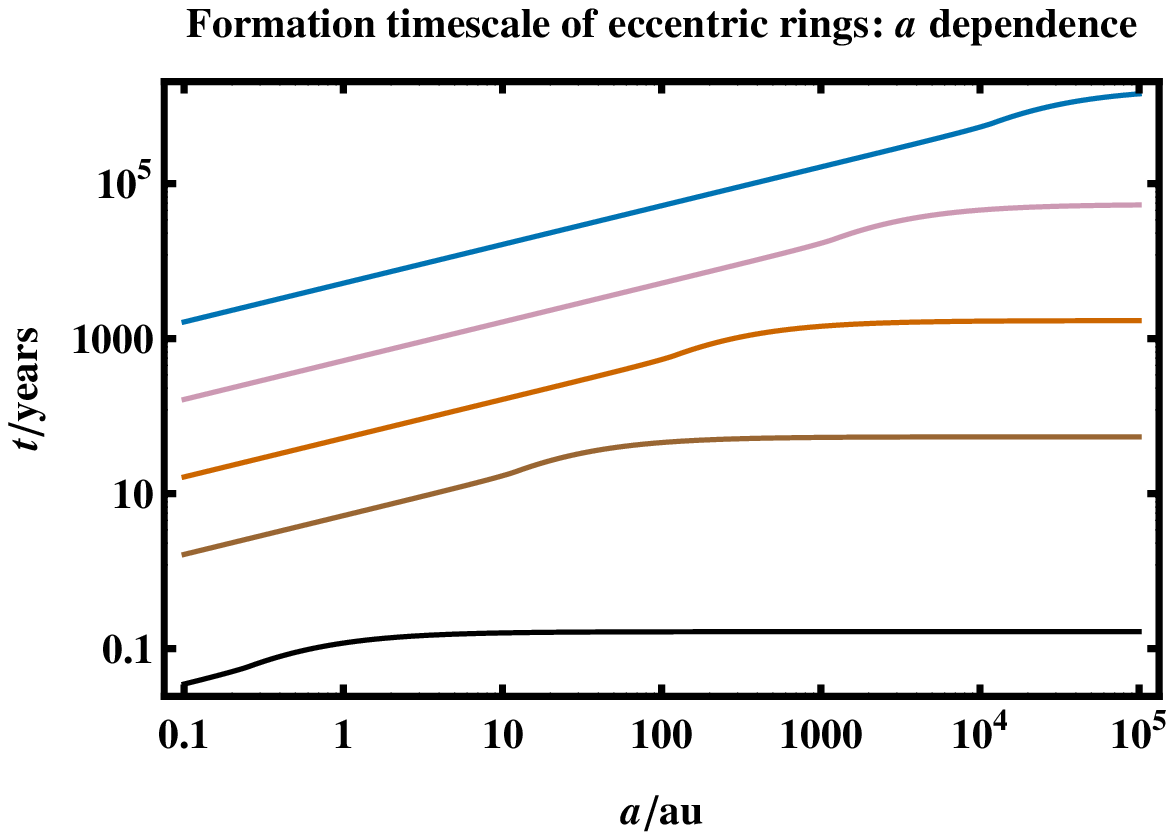,height=6.5cm,width=8cm}
}
\caption{
Time which debris takes to fill an eccentric thin ring after
an instantaneous disruption of an asteroid at a distance $r_{\rm b}$
from a WD of mass $0.6 M_{\odot}$. The asteroid's
original radius and original orbital semimajor 
axis are $R$ and $a$, and $r_{\rm c} = 0.017$ au.  The top and 
bottom panels express filling time in
terms of orbital period $T$ and in years, respectively.  The plots
demonstrate that the formation timescale is highly dependent
on $r_{\rm b}$, $R$ and $a$.
}
\label{spread1}
\end{figure}

\begin{figure}
\centerline{
\psfig{figure=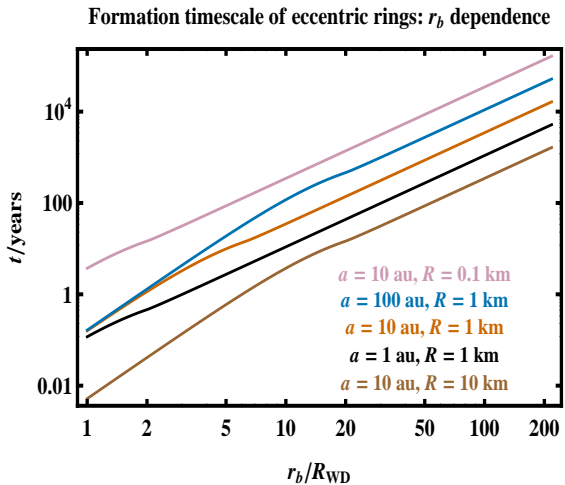,height=6.5cm,width=8cm}
}
\caption{
Like Fig. \ref{spread1}, except highlighting the dependence
on the disruption distance.  For all combinations presented here,
when disruption occurs within $10 R_{\rm WD}$, the ring
will form within about 100 yrs. 
}
\label{spread2}
\end{figure}

\begin{figure}
\centerline{
\psfig{figure=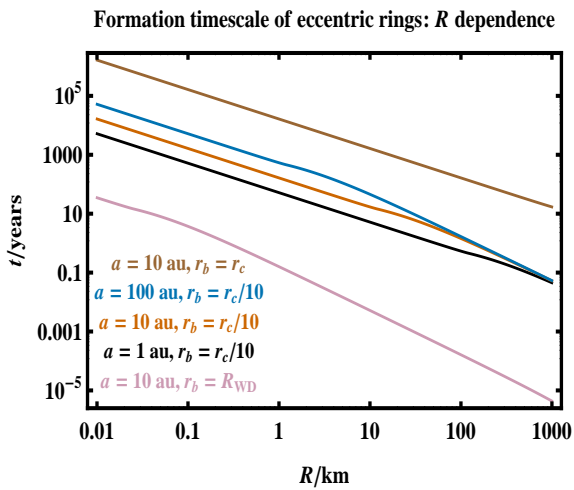,height=6.5cm,width=8cm}
}
\caption{
Like Fig. \ref{spread1}, except highlighting the dependence
on the radius of the asteroid.  The plot shows that any asteroids 
with $R > 1$ km which skim the WD surface will fill out a ring 
within a couple months.  Also, this plot demonstrates that the
filling time has a weak dependence on the semimajor axis of the
orbit for constant $R$.
}
\label{spread3}
\end{figure}

This formula (equation \ref{fill1}) allow us to generate eccentric disc 
formation timescales purely analytically.  The results are presented in Figs. 
\ref{spread1}-\ref{spread3} for our fiducial $0.6 M_{\rm WD}$ WD with $r_{\rm c} = 0.017$ au
(see equation \ref{maxrc1}).
Figure \ref{spread1} illustrates the timescale in terms of years (upper panel) 
and original orbital periods (lower panel) for five different combinations of 
the disruption location $r_{\rm b}$ and the original asteroid radius $R$ as a function of $a$.
The top (blue) curves represent the maximum possible filling time for a $1$ km 
asteroid, which is several orders of magnitude less than a WD cooling time of 1 Gyr.
At the other extreme, disruptions where the asteroid skims the WD surface will 
fill out an eccentric ring with debris in a couple months.  Note that the curves 
in the bottom panel level out beyond a 
particular semimajor axis value, one that increases with disruption location 
and decreases with asteroid radius.

Figures \ref{spread2} and \ref{spread3} instead highlight the dependence
on the disruption distance as a function of both $R_{\rm WD}$ and $R$, 
by placing those values on the $x$-axes.  Figure \ref{spread2}
suggests that any asteroids thrown in from an exo-asteroid belt (at $\approx 5$ au)
or an exo-Kuiper belt (at $\approx 30$ au) which reach pericentre values
within $10R_{\rm WD}$ will fill out an eccentric ring within about 100 yrs.
Figure \ref{spread3} further illustrates that the formation timescale is also
strongly dependent on the asteroid's radius.  Here the $x$-axis extends to
values of $1000$ km, which is roughly twice the value of the radius of the
largest known asteroid (Ceres) and is comparable to that of small 
planets\footnote{\cite{veretal2013} and \cite{musetal2014}  
specifically considered how dynamical instabilities in multi-planet
systems may cause a collision with a WD and a planet.}. 

As previously mentioned, the results in this section are dependent on the assumption
that the breakup is instantaneous and thorough, such that post-breakup, all particles 
will evolve independently of one another.  Our numerical simulations, which are reported in the
next section, show that the breakup is never strictly instantaneous.  Rather, clumps
of particles remain bound for more than one pericentre passage.  As the pericentre
of the orbit increases, our assumptions break down further, as the clumps become larger
and are more strongly bound.  Hence, the formulae here are best-suited for close pericentre
passages.  Further, because our final formula (equation \ref{fill1}) is 
independent of particle mass or size, the formula should be applicable to asteroids with different 
particle size distributions as long as the extent of clumping for these rubble piles is negligible.
These distributions may be significantly influenced, or even primarily determined,
by destructive processes occurring during the star's giant branch evolutionary
phases \citep{veretal2014b}.

%%%%%%%%%%%%%%%% Figure
%%%%%%%%%%%%%%%% Figure

%%%%%%%%%%%%%%%% Figure
%%%%%%%%%%%%%%%% Figure

%%%%%%%%%%%%%%%% Figure
%%%%%%%%%%%%%%%% Figure

\section{Numerical simulations}

Now we complement our theoretical predictions with 
numerical simulations of a rubble-pile asteroid.  Here we describe the 
code used, the internal structure of the rubble piles that are modelled, 
the timestep adopted, and finally our simulation results. 
Our discussion of the timestep adopted may be widely applicable
to other similar $N$-body codes.

\subsection{Numerical disruption code}

We use a modified version of the well-established $N$-body gravity tree 
code {\tt PKDGRAV} \citep{stadel2001}.  The major modification is the ability 
to detect and resolve collisions \citep{ricetal2000}.
For our simulations, the $N$ bodies are equal-mass and equal-radius particles 
which initially comprise a single gravitational aggregate known as a 
{\it rubble pile} (although the code is flexible enough to handle interactions between
multiple rubble piles; see \citealt{leietal2000}).
As the rubble pile becomes disrupted, the particles' motion is consistently treated 
by the code.  The WD, or any parent
star, may be introduced into the code, but is not treated as one of the $N$ bodies.
Instead, the WD is treated as a gravitational point mass (with zero radius).  
Hence employing a realistic mass-radius relation (equation \ref{MR})
in the setup is crucial so that orbits do not pass through a region where the WD 
should reside.

The integrator used is a second-order leapfrog integrator, which is symplectic in the absence 
of collisions.  For a more extensive discussion on the properties of this integrator, 
see \cite{ricetal2000}.

\subsection{Rubble pile characteristics}

\begin{figure}
\includegraphics[width=8cm]{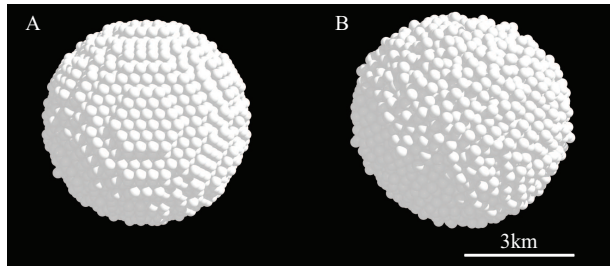}
%\begin{figure}
%\centerline{
%\psfig{figure=rubble.eps}
%}
\caption{
Rubble-pile asteroids, each composed of about 5000 indestructible hard spheres that we denote as
{\it particles}.  Asteroid A is hexagonally-packed, and asteroid B is randomly-packed.  
Disruption is very weakly dependent of the packing method as long as the asteroid is roughly spherical.
}
\label{rubblepiles}
\end{figure}

The number of particles and their bulk shape might significantly affect the details
of disruption \citep[e.g.][]{ricetal1998}.  Here we consider only roughly spherical
rubble piles of 5000 particles, with two different internal structures.  One structure consists
of hexagonally-packed particles, and the other randomly-packed particles
(see Fig. \ref{rubblepiles}).  We find, in concert with previous studies which use {\tt PKDGRAV}, that our choice
of rubble-pile configuration makes no discernibly important difference in the outcome
of our simulations.  Consequently, we henceforth report results from only our randomly-packed
rubble-pile simulations.  We adopt an asteroid mass of about $2.26 \times 10^{14}$ kg.
The semi-axes of the rubble pile are about 3.25 km, 3.05 km and 2.99 km, yielding a
bulk density of about 1.82 g/cm$^3$. The spins of all of the particles are randomly oriented.

%% hex: > convert ss.000200.ras -crop '2000x1700+1000+1150' -negate -filter Hermite ss.000200.gif
%% ran: > convert ss.002260.ras -crop '1700x1550+100+375' -negate -filter Hermite ss.002260.jpeg

%%\centerline{
%%\psfig{figure=hex000200fromtest4.eps,height=5.1cm,width=6cm}
%%}
%%\centerline{
%%\psfig{figure=ran002260fromtest6.eps,height=5.4705882353cm,width=6cm}
%%}
%%%%%%%%%%%%%%%% Figure
%%%%%%%%%%%%%%%% Figure

\subsection{Timesteps}

We use a fixed timestep in our simulations.  Determining the appropriate value of this timestep
is crucial to ensure numerical convergence and accurate results.  To guide our intuition for
the correct value to adopt, we take note of five applicable timescales.  The first is the 
{\it dynamical} timescale

\begin{equation}
\tau_{\rm dyn} \propto \frac{1}{\sqrt{G\rho}}
.
\label{tdyn}
\end{equation}    

\noindent{}For asteroids, $\tau_{\rm dyn} \sim 1$ hour. Previous investigations using 
{\tt PKDGRAV} \citep[e.g.][]{leiric2002} demonstrate that adopting a timestep
$\Delta t$ which is about two orders of magnitude smaller than $\tau_{\rm dyn}$ 
($\Delta t \approx 50$s) sufficiently resolves the collisions amongst the particles 
in a rubble pile.

%%%%%%%%%%%%%%%% Figure
\begin{figure}
\centerline{
\psfig{figure=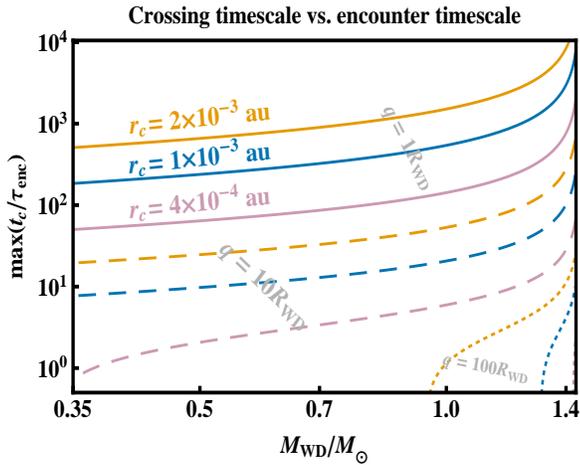,height=6.5cm,width=8cm}
}
\caption{
Demonstration that timestep sampling within the disruption sphere is independent
of orbital timestep sampling.  Ensuring adequate sampling in the former will
not guarantee adequate sampling of the latter.  The
effect is particularly pronounced for pericentres within a few WD radii, and is
a strong function of $r_{\rm c}$.  The solid, dashed and dotted lines refer to
$q=1,10,100 R_{\rm WD}$.  Visual changes of these curves when sampling different $a$ 
values greater than 1 au are imperceptible.
}
\label{timeratio}
\end{figure}
%%%%%%%%%%%%%%%% Figure

%%%%%%%%%%%%%%%% Figure
\begin{figure}
\centerline{
\psfig{figure=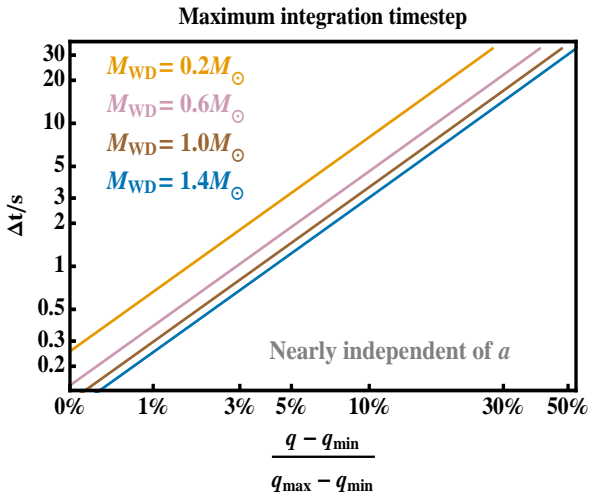,height=6.5cm,width=8cm}
}
\caption{
Maximum numerical integration timestep $\Delta t$ as a function of pericentric distance.
Values of $q_{\rm min}$ and $q_{\rm max}$ are given by equation (\ref{qmaxqmin}).
This plot illustrates that for the most relevant disruption region, within
the inner half of the disruption sphere, the required
timestep is always less than the dynamical timestep of 50 s.
}
\label{tsfig}
\end{figure}
%%%%%%%%%%%%%%%% Figure

The second timescale is the {\it orbital} timescale

\begin{equation}
\tau_{\rm orb} = \frac{2\pi a^{3/2}}{\sqrt{GM_{\rm WD}}}
.
\label{torb}
\end{equation}

\noindent{}The orbital timescale of every known asteroid, comet or planet exceeds 1 hour by several orders of magnitude.  Hence, typically, 
$\tau_{\rm orb} \gg \tau_{\rm dyn}$.  In symplectic simulations of point mass planets orbiting a star, a well-utilised
rule of thumb is $\Delta t \le (1/20) \tau_{\rm orb}$ \citep{dunetal1998}.

The third timescale is simply the disruption sphere crossing timescale, $t_{\rm c}$.  
We must ensure that the rubble pile is sufficiently sampled within the disruption sphere.
So either the disruption sphere crossing timescale or the dynamical timescale
dictates the limiting timestep.  However, there is one more consideration.

The fourth timescale is the {\it encounter} timescale

\begin{eqnarray}
\tau_{\rm enc} &\approx& \frac{q}{v_q}
= \sqrt{\frac{q^3}{G M_{\rm WD}}} \left(2 - \frac{q}{a} \right)^{-1/2}
\label{tenc1}
\\
 &\approx& 
205 {\rm s} \left( \frac{M_{\rm WD}}{0.6 M_{\odot}} \right)^{-1/2}
\left( \frac{q}{0.001 \ {\rm au}} \right) \frac{1}{\sqrt{1+e}}
,
\label{tenc2}
\end{eqnarray}

\noindent{}which is the timescale for gravitational interaction at the
closest approach distance.

Except near the edge of the disruption sphere, $\tau_{\rm enc} < t_{\rm c}$.
However, we can obtain a more meaningful comparison by taking the ratio of these
two timescales.  The value of $\left(t_{\rm c}/\tau_{\rm enc}\right)$ is well-approximated at all
relevant values of $a$ by

\begin{equation}
\lim_{a \rightarrow \infty}{\left(\frac{t_{\rm c}}{\tau_{\rm enc}}\right)}
=
\frac{4}{3}
\left[
\frac
{r_{c}^2 + q r_c - 2q^2}
{q^{3/2}\sqrt{r_c - q}}
\right]
\label{limtcqcdiff}
\end{equation}

\noindent{}which monotonically decreases as $q$ shifts from $R_{\rm WD}$ to $r_{\rm c}$, and hence takes on a
maximum value at $q = R_{\rm WD}$.  This ratio is plotted in Fig. \ref{timeratio}.  The figure 
demonstrates that the timestep restrictions near
the WD surface are demanding.  
Adopting the dynamical timescale constraint of $\Delta t \approx 50$ s will fail to sufficiently 
resolve the encounter within a few $R_{\rm WD}$ for the lowest-mass WDs, and within tens of $R_{\rm WD}$ for 
the highest-mass WDs. 

The fifth timescale is the {\it collision} timescale, $\tau_{\rm col}$, which represents the ratio
of a characteristic interparticle distance to the relative velocities of the particles.  Because the
minimum possible size of a particle orbit is the diameter of the WD, and the particles 
orbit in the same direction around the WD after disruption, for our purposes 
$\tau_{\rm col} > \tau_{\rm dyn}$ always.

%%%%%%%%%%%%%%%% Figure
\begin{figure*}
\centerline{
\psfig{figure=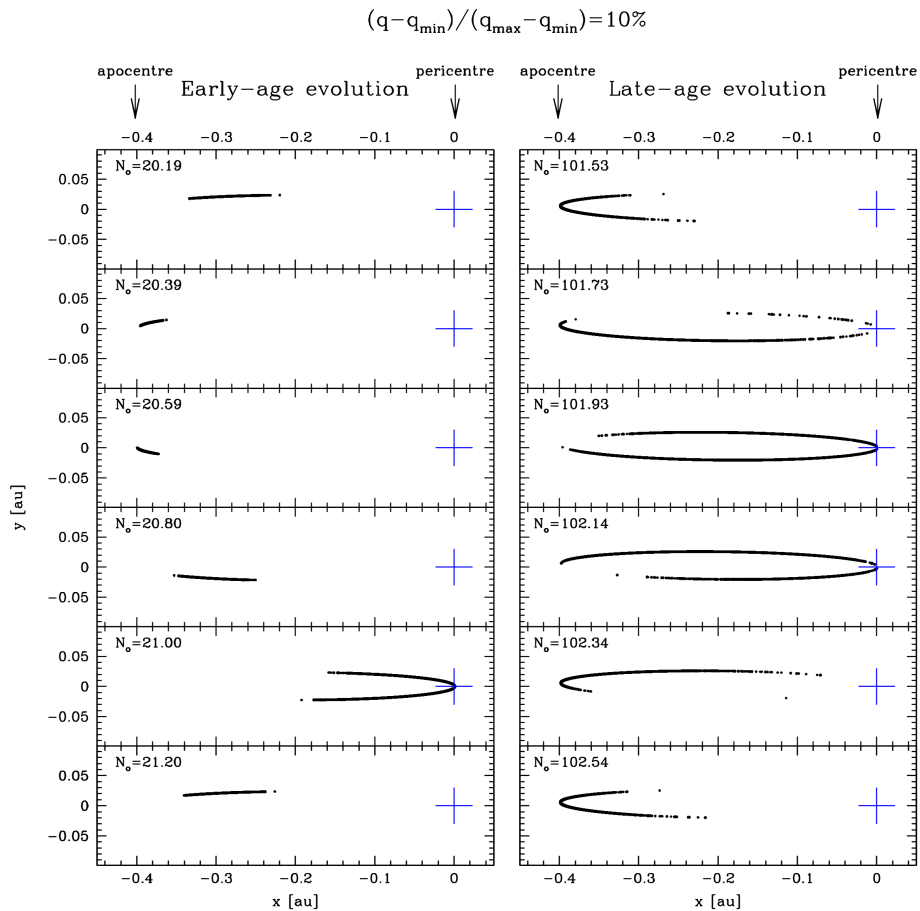,width=13cm}
}
\caption{
The evolution of a disrupted rubble pile after $N_{\rm o}$ orbits when
$N_{\rm o} \approx 20$ ({\it left panels}) and $N_{\rm o} \approx 100$ ({\it right panels})
from the {\tt PKDGRAV} numerical code.  Snapshots in equal time 
increments of about 0.202 orbits are displayed from top to bottom.
The motion is counterclockwise around the WD, which is denoted with a blue
cross and set at the origin. Although the simulations here have $a = 0.2$ au, {\it 
the qualitative evolution is self-similar for greater orbital distances because 
the disruption characteristics are largely independent of semimajor axis for $a \gtrsim 0.1$ au}.
In reality, $a > 1$ au; the value of $a = 0.2$ au was chosen entirely for computational reasons.   
The plot illustrates the speed at which an eccentric debris ring fills out
when the orbital pericentre $q$ satisfies 
$\left(q - q_{\rm min}\right)/\left(q_{\rm max} - q_{\rm min}\right) = 10\%$.
}
\label{disrupt10}
\end{figure*}
%%%%%%%%%%%%%%%% Figure

%%%%%%%%%%%%%%%% Figure
\begin{figure*}
\centerline{
\psfig{figure=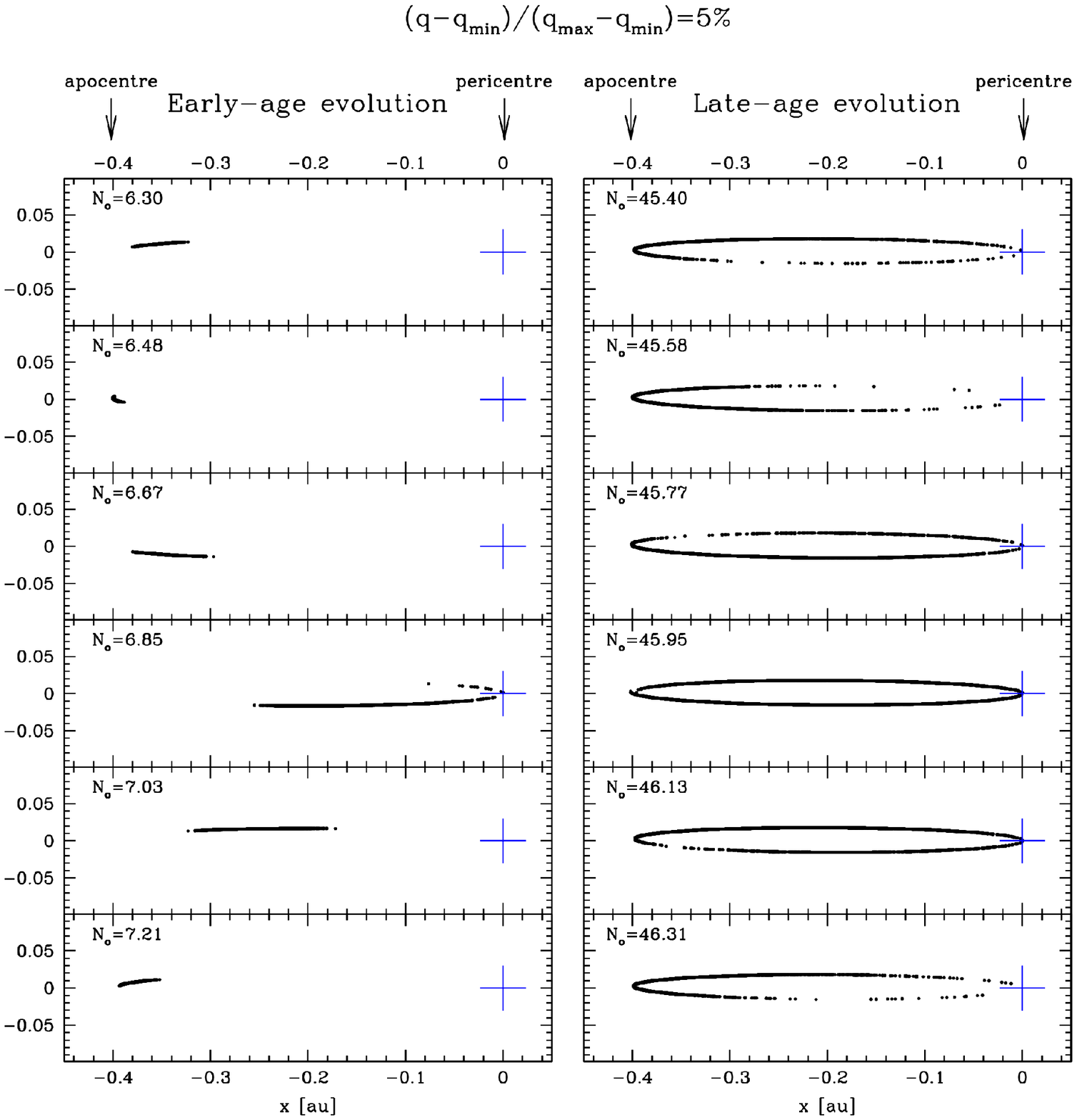,width=13cm}
}
\caption{
Similar to Fig. \ref{disrupt10}, except for 
$\left(q - q_{\rm min}\right)/\left(q_{\rm max} - q_{\rm min}\right) = 5\%$.
Here, the left and right panels show the disrupted rubble pile 
for $N_{\rm o} \approx 6$ and $N_{\rm o} \approx 45$, respectively, demonstrating
that the rings fill out more quickly than in Fig. \ref{disrupt10}.  Snapshots 
in equal time increments of about 0.183 orbits are displayed from top to bottom.
}
\label{disrupt05}
\end{figure*}
%%%%%%%%%%%%%%%% Figure

Finally, these considerations lead us to adopt the following timescale for each of our simulations

\begin{equation}
\Delta t = {\rm min} \left(50 {\rm s}, \ \frac{t_{\rm c}}{20}, \ \frac{\tau_{\rm enc}}{30} \right)
.
\label{timestep}
\end{equation}

\noindent{}The factor of 30 in the last denominator arises from our preliminary simulation
suite.  We discovered that for higher timesteps, the accumulated error over tens of orbits
noticeably alters the argument of pericentre of the orbit.

Because the value of $\Delta t$ crucially affects the CPU running time of our simulations, we
now quantify how $\Delta t$ varies with $a$ and $q$.  Figure \ref{tsfig} illustrates this
dependence, where we have used equations (\ref{newtc}) and (\ref{tenc1}) to compute $t_{\rm c}$ and
$\tau_{\rm enc}$, with $r_{\rm c}$ given by equation (\ref{maxrc1}).  Although its dependence on $a$ 
is negligible and dependence on $M_{\rm WD}$ is weak, $\Delta t$ is less than the dynamical 
timescale of 50s for over half of the distance from $q_{\rm min}$ to $q_{\rm max}$.

\subsection{Simulation results}

We seek to answer three important questions from our simulation results: (1) Do rubble piles 
actually form highly-eccentric rings, as has been theorised?  (2) If so, what is the 
collisional nature of the rings?  (3) What is the timescale to fill out the rings with 
rubble?  

\subsubsection{Qualitative answers to these questions}

Our simulations show that highly-eccentric structures are indeed formed, and are 
filled-in over time in the rough shape of a ring.  Deviations from a perfect filled ring are due 
to the chaotic nature of the non-instantaneous dissociation of thousands of 
mutually-interacting particles
and the amount of material (or number of particles) inside of the asteroid.  These deviations
take the form of arcs which are void of particles, and brakes in the shape of the 
annulus itself.  The distortion of the shape becomes pronounced only within a few WD
radii.

Importantly, our disrupted asteroid eventually completely dissociates in all cases.
Accordingly, each particle eventually orbits the WD without being influenced by
any other particle.  The result is a {\it collisionless} collection of particles,
each of which propagate according to the classic two-body problem.  
Gravity alone cannot cause accretion.
{\it They will never accrete onto the WD unless influenced by other forces}.   
The addition of other forces, such as radiation from the WD
and non-gravitational forces from sublimation, are topics for future work.  
The timescale at which complete dissociation occurs
is a strong function of the initial conditions and particularly the pericentre.  Further,
complete dissociation is a conservative notion.  We have found that well before this
condition is satisfied, the particles are effectively collisionless, with just a few
2-particle clumps hanging on for the same amount of time that is taken for the rest of the
asteroid to dissociate.

\subsubsection{Simulation details}

Our simulations were carefully chosen to both showcase important behaviour and finish
running on plausible timescales ($\sim 1$ month).  The duration of the simulations 
is severely limited by a timestep which is tiny (see Fig. 8) compared to typical numerical 
simulations of planetary systems.  Consequently, any simulations with semimajor axes of more than a 
few tenths of an au and a pericentre beyond a few WD radii would require over one month 
in real time to model a single orbit with {\tt PKDGRAV}.  
Fortuitously, time spent in the disruption sphere is very weakly dependent on semimajor 
axis (Fig. \ref{tc1}), allowing us to adopt $a = 0.2$ au and hence model tens of orbits 
self-consistently with our code.  The choice of 0.2 au is motivated only by computational limitations.
In reality, no asteroid should exist in a WD system on a $a = 0.2$ au orbit. Rather, asteroids
should harbour semimajor axes greater, or much greater, than 1 au, but fortunately, the problem scales 
extremely well for semi-major axes greater than 0.1 au (e.g. Fig. \ref{tc1}).

We display results from two of our simulations in the form of snapshots
in Figs. \ref{disrupt10}-\ref{disrupt05}.  These simulations
have values of $(q - q_{\rm min})/(q_{\rm max} - q_{\rm min})$ of 10\% and 5\%
respectively.  
The constant timesteps adopted for 
the simulations were about $7.356$ and $2.772$ seconds,
respectively, in close accordance with Fig. \ref{tsfig}.  Hence,
the number of steps required to complete one original orbital
period were approximately 495,000, and 1,314,000.
The original eccentricities of the orbits are about 0.9934
and 0.9966.  The simulations assume a WD mass
of $0.6 M_{\odot}$.  The asteroids all begin their motion at 
$\Pi_0 = -48.96^{\circ}$, a value which affords a ``lead-in'' time of $t_c$
to the disruption sphere, where $t_c$ is computed according to
the orbit which skims the WD. 

Figures \ref{disrupt10}-\ref{disrupt05} illustrate how disruption typically forms an arc of
material which gradually expands into a ring.  The expansion is due to the velocity gradient
of the particles.  These velocities are determined by their last combined interaction with both
the WD and another particle just before dissociation from that particle.  This process does
not reproduce a continuous and uniform velocity distribution because the disruption is not
instantaneous.  Nevertheless, the approximation used in Section 4 and Figs. 
\ref{spread1}-\ref{spread3} correspond well with the simulation results: for $a = 0.2$ au,
and assuming that the disruption of a roughly 3 km-radius asteroid occurs at the pericentre, the fill-out
time is expected to be a few tens to a couple hundred orbits.  For semimajor axes of a few au,
the fill-out time would then correspond to tens or hundreds of years.

By focusing on the middle of the 
figures, one can visually discern a slight artificial precession of the ellipse due to accumulated
numerical error (despite our conservative timesteps).  This precession, which is not due to the general relativistic precession (equation \ref{GRinf}) is about twice as prominent in the right column of Fig. \ref{disrupt10} than in Fig. \ref{disrupt05} partly due to the former being run for about twice as many orbits.

Finally, in order to test the robustness of our results against the resolution
of our rubble piles, we have performed additional simulations with rubble piles
which contain nearly 1,000 and 10,000 particles.  In each case, we performed
simulations with $(q - q_{\rm min})/(q_{\rm max} - q_{\rm min})$ values 
of 10\% and 5\%.  We find that like in the 5,000-particle case, (1) highly-eccentric 
collisionless rings are formed, and (2) greater resolution (number of particles) 
improves the homogeneity of the resulting ring that is formed.  More particles
help fill in gaps in the ring.  We have also repeated our 1,000-particle case
using both of the above $q$ values but a different tangential coefficient of
restitution (0.5, instead of 1.0).  The results were qualitatively similar.

Recent work featuring a higher level of sophistication in the modelling of tidal 
disruption \citep[e.g.][]{movetal2012,yuetal2014} showcases potential future directions for follow-up
studies.  In these cases, particles are idealised not as indestructible hard
spheres, but rather as soft spheres \citep{schetal2012}.  In the soft sphere discrete element
method, rolling and twisting friction may be incorporated between particles, and 
particles may share multiple points of contact.  Particles also need not be modelled as spheres;
\citep{movetal2012} instead use irregular, polyhedral grains.

\section{Summary}

We have investigated an important step in the process of polluting WDs with circumstellar 
material: the tidal disruption of bound asteroids which veer into the WD's Roche radius.  
We conclude that while an initially spherical asteroid perturbed onto an eccentric 
orbit may be tidally disrupted by 
a WD to form a highly eccentric ring of debris, this ring is collisionless without
the influences of additional perturbative forces.  Without these forces, the disrupted asteroid will 
not accrete onto the WD, importantly demonstrating that gravity alone is insufficient to 
produce WD pollution.  These results motivate future investigations which would detail how 
eccentric collisionless rings can form a close-in circumstellar disc 
(with the approximate dimension of the disruption radius), from where the debris eventually accretes
onto the WD.

Although this paper considered the disruption of just a single asteroid, multiple asteroids could
arrive at the WD's disruption radius in quick succession, as the co-orbital 
fragments of comet Shoemaker-Levy 9 did at Jupiter.  Consequently, the filling time for a ring 
will decrease.  The resulting filling timescale depends upon the number of tidally-disupted
asteroids and the relative orientations of the incoming objects, but not their masses (see discussion
after equation \ref{fill1}). The masses will determine the extent and number of the gaps 
in the ring, causing it to appear as a series of loosely- or strongly-connected arcs.  
If the incoming objects are
not initially co-orbital, then they will be disrupted at different pericentres, and instead
form a series of rings akin to the ring system of a giant outer Solar system planet.  A packed
collection of rings may be classified as a disc.

Our more specific findings include a characterisation of the interplay between the extremely eccentric 
orbits of these asteroids and the WD's Roche radius 
(Section 3).  Consequently, we conjecture that the characteristics of disruption is largely independent of 
semimajor axis (Fig. \ref{tc1}), and highly dependent on the pericentre (Fig. \ref{tc2} and equation 
\ref{limtc}).  Our work has revealed that the debris follows the original orbit, first as a short arc and 
then later as a full ring 
after a time given by equation (\ref{fill1}) and Figs. \ref{spread1}-\ref{spread3}. Numerical 
simulations with the rubble-pile integrator {\tt PKDGRAV} disclose that the debris does not uniformly fill 
out the ring (Figs. \ref{disrupt10}-\ref{disrupt05}).  To prevent significant artificial precession due to 
accumulated numerical error, the required maximum timesteps for these types of 
simulations are extreme, often on the order of one second (equation \ref{timestep}).

\section*{Acknowledgments}

We thank the referee for helpful suggestions.
We also thank Mia Mace for ray-tracing Fig. \ref{rubblepiles}, and J.J. Hermes, Derek C. Richardson, 
and Steinn Sigurdsson for useful discussions. The research leading to these results 
has received funding from the European Research Council under the European Union's 
Seventh Framework Programme (FP/2007-2013) / ERC Grant Agreement n. 320964 (WDTracer).

\label{lastpage}
\end{document}